**PREPRINT DEPARTMENT OF PHYSICS AND ELECTRONICS,UNIVERSITY OF PUERTO RICO AT HUMACAO**

# EFFECTS OF NON-SPHERICAL SYMMETRY ON BINARY ORBITS OF ASTEROIDS AND COMETS

**W. BRUCKMAN, J. C. CERSOSIMO, L. ROSA**

## ABSTRACT

We develop a theoretical framework for the calculation of orbits for a system consisting of a spherical object and a non-spherical body, which is then specialized to a prolate ellipsoid. Particular trajectories are presented that illustrate a drastic contrast between the familiar elliptical orbits of spherical binary systems and the trajectories around the prolate spheroid. We also show here, and in a media video representation of the computed orbits, how the spherical satellite instantaneous orbital plane and eccentricity evolve. We also explicitly verify the conservation of the total angular momentum and energy of the system, prolate plus satellite, while the intrinsic rotational angular momentum and energy of the prolate changes with time at the expense of the orbital energy and angular momentum of the sphere. We then consider a particular orbit where an initially bound satellite gains sufficient orbital energy and eventually escapes, with its total energy now positive. The inverse process, where a satellite is captured by a prolate, is also considered, and we determine the probability of this event occurring, as a function of the initial relative velocity and parameter of impact of the system. We end with a discussion of a plausible scenario where an escaping satellite in the Oort cloud could wind up with a new heliocentric Earth`s crossing orbit. In the Appendices we develop the necessary equations for the application of the above formalism to orbits around a general homogeneous ellipsoid.

## 1. Introduction And Summary

Satellites of asteroids and comets are not uncommon(see for example Margot et al(2002). Therefore, the orbits of satellites around a non spherical body is an interesting topic for study. In this work we discuss the problem of computation of orbits for binary systems consisting of a spherically symmetric body, $m_1$ , and a body, $m_2$ , of arbitrary symmetry, which is then specialized to a prolate homogeneous ellipsoid.

There are more technical publications that deal with the applications of the gravitational two-body problem to specific realistic astrophysical systems. These numerical experiments are essential steps for the understanding of the observed binary systems. However, working with real configurations requires the use of advanced and complicated procedures, which provide good approximations to the observed phenomena, but are not suitable to didactic expositions. Alternatively, we can treat idealistic configurations with exact equations of motions in order to facilitate the pedagogical and heuristic presentation of the physics involved. The prolate spheroid is chosen as our non-spherical body since this form is a good approximation to elongated bodies such as asteroids and comets; furthermore, their gravitational potentials and fields are well known analytically. On the other hand, the studies of orbits around systems with other symmetries will only require a straightforward modification of the intervening forces in our computations and thus, we also develop in the Appendices the potentials and fields external to general homogeneous ellipsoids .

In section 2 we develop the necessary equations of motions, where Newton`s second law is combined with the conservation law for total angular momentum of the system, to provide the differential equation for the dynamics of the six degrees of freedom of the configuration. The three equations for the conservation of angular momentum are used because they are of first order in the time derivative of the linear and angular coordinates, and therefore they are simpler and more convenient than the three second order Euler-Lagrange equations involving the angular accelerations of the prolate.

An example of a typical trajectory, lasting about 47 days, is shown in section 3. The whole orbit is subdivided into three segments, each one with a period of about one day, separated from each other by intervals of about 22 days. In these figures it is clearly illustrated that the plane of the “instantaneous” orbit is significantly changing over intervals of weeks, in contrast to the constant orientation of the orbital plane for spherical bodies in binary systems.

An illustration of an orbit where the two bodies have comparable mass is presented in section 4 (see also http://www.uprh.edu/fisica/orbita.html). The pictures show how the rotation of the proloid is strongly affected by the spherical companion, since a substantial interchange of angular momentum and energy is occurring between the prolate intrinsic rotation and the orbital degrees of freedom of the system. Thus, the spherical mass could gain (or lose) kinetic energy at the expense of the rotational energy of the prolate. However, we still have the conservation of total energy and total angular momentum of the system, prolate plus spherical body, which is illustrated explicitly in this section. It is also shown that the intrinsic rotational energy and angular momentum of the prolate is changing as a function of time.

In the last section, we consider a particular orbit that demonstrates how an initially bound satellite gains kinetic energy until it eventually reaches an escape velocity and the orbit becomes open. Once the satellite escapes, its orbit around the Sun will differ from the orbit of the original binary system center of mass. Then, it is natural to ask under what conditions this new satellite trajectory could become a collision treat to Earth because of a drastic change in its perihelion. We find, in section 5, that this situation is plausible for binary comets in the Oort cloud.

The fact that satellites can escape from bound orbits tells us that an initially free body can also have a probability to be trapped by another. We can see this explicitly, by running backwards in time the orbit of the escaping satellite, discussed about. We then calculate the probability of such events, for a specific prolate and a spherical mass encounter. It turns out that this probability is a remarkably simple function of the initial relative speed of the bodies and the parameter of impact of the system.

## 2. The Two Body Problem

From Newton´s second law it follows that the acceleration of the center of mass of a body, $m_{1,}$ relative to the center of mass of a second body, $m_2$, is given by

$$\frac{d^2\vec{r}}{d^2t} = \frac{\vec{F}}{\mu} \quad , \qquad (1)$$

where $\vec{r}(x, y, z)$ is the vector position of the center of mass of $m_1$ relative to an origin at the center of mass of body $m_2$, $\mu = m_1 m_2/(m_1 + m_2)$ is the reduced mass, and $\vec{F}$ is the gravitational force on $m_1$ produced by the interaction with $m_2$. In general, $\vec{F}$ would be a function of $(x, y, z)$ and the orientation in space of the two bodies. However, in our situation, $m_1$ is spherically symmetric, and therefore only the rotational degrees of freedom of $m_2$ counts. This orientation can be conveniently described by the well known set of three Eulerian angles $(\phi, \theta, \psi)$ illustrated in Figure (1), where the coordinate system $(x', y', z')$ axe are fixed in $m_2$, while $(x, y, z)$ axe are fixed in space. Thus, we have, writing Eq. (1) in component form,

$$\frac{dx_i{}^2}{dt^2} = \frac{1}{\mu} F_{x_i}(x, y, z,\ \psi, \theta, \emptyset);\ \ x_1 = x,\ x_2 = y,\ x_3 = z \qquad (2)$$

Since the above three differential equations include six unknowns, we need three additional equations to be able to integrate the equations of motion and find the orbits. We will furnish these additional equations using the conservation law for total angular momentum, which is valid for isolated systems. For a two-body problem where one of them is a sphere, one can write without loss of generality,

$$\vec{L} = \mu\vec{r}x\frac{d\vec{r}}{dt} + I.\vec{\omega} = \overrightarrow{L_0} = Constant, \tag{3}$$

where the first term represents the orbital angular momentum of the system, and $\vec{l} \equiv I.\vec{\omega}$ is the intrinsic angular momentum of $m_2$ . $I$ is the tensor of inertia, $\vec{\omega}$ the angular velocity vector in the $(x, y, z)$ coordinate system, and we have ignored the intrinsic angular momentum of $m_1$ since for spherically symmetric bodies, this is a constant of the motion.

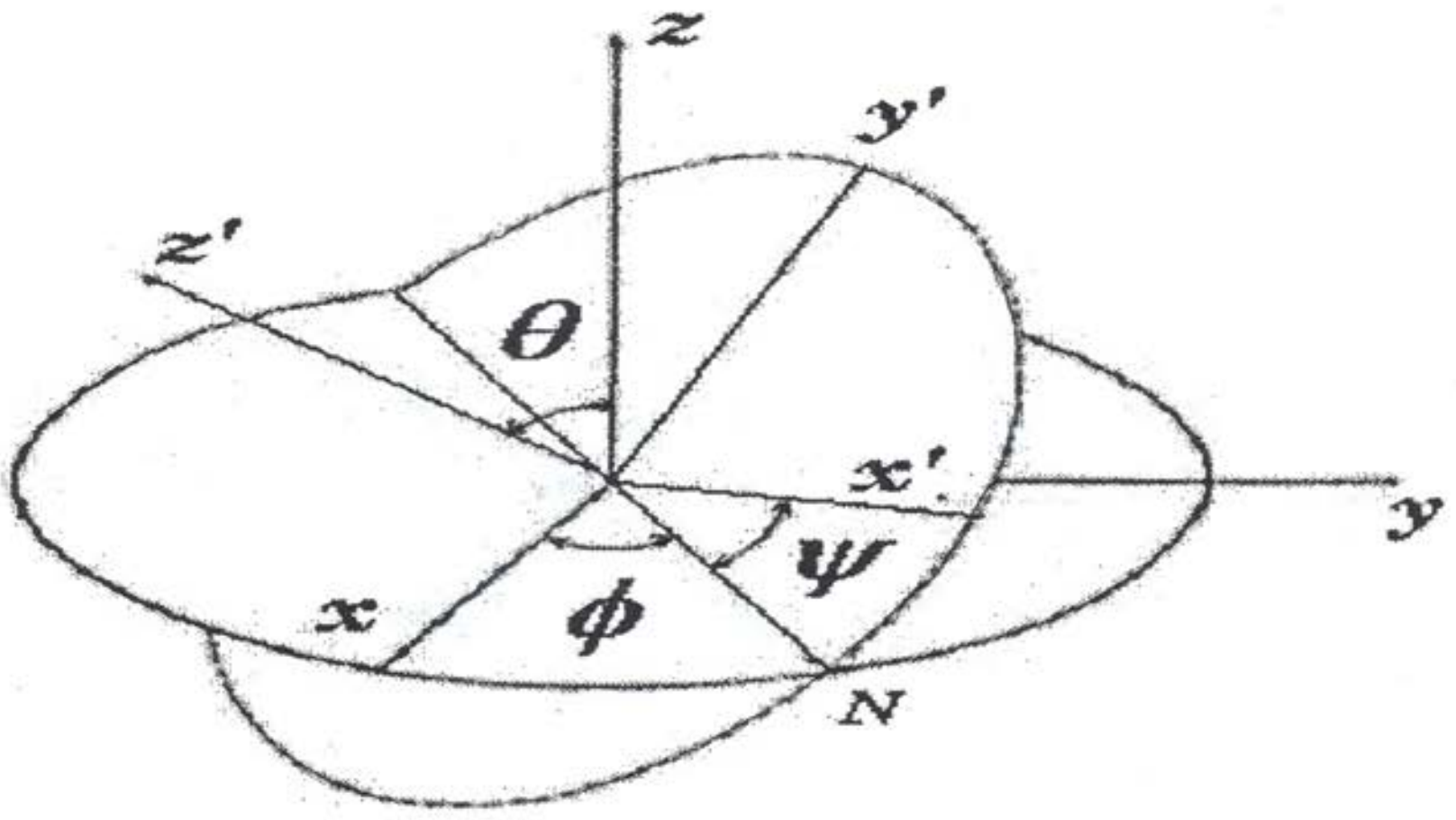

**Figure 1**: The Euler angles

It is possible to choose the coordinate axes $(x', y', z')$ in the body such that the tensor of inertia is reduced to a diagonal form. These directions are called “the principal axe of inertia”, and the corresponding values of the diagonal components of the tensor are called “the principal moments of inertia”: $I_{x`x`}, I_{y`y`}, I_{z`z`}$ . Thus, we have

$$I^{'} = \begin{bmatrix} I_{x'x'} & 0 & 0 \\ 0 & I_{y'y'} & 0 \\ 0 & 0 & I_{z'z'} \end{bmatrix}. \tag{4}$$

The intrinsic angular momentum in the principal axes of inertia $\vec{l}` = I`.\vec{\omega}`$ is related to $l$ by the transformations

$$\vec{l}` = A\vec{l} \tag{5}$$

$$\vec{l} = A^{-1}\,\vec{l}` = A^{-1}\,I`.\vec{\omega}` \tag{6}$$

where $A$ is the Euler matrix, which transforms vector components in the $x, y, z$ coordinate system to vector components in the $x', y', z'$ system:

$$\begin{pmatrix} x` \\ y` \\ z` \end{pmatrix} = A \begin{pmatrix} x \\ y \\ z \end{pmatrix}, \tag{7}$$

and is given by Goldstein(1980):

$$A = \begin{pmatrix} \cos\psi\cos\phi - \cos\theta\sin\phi\sin\psi & \cos\psi\sin\phi + \cos\theta\cos\phi\sin\psi & \sin\psi\sin\theta \\ -\sin\psi\cos\phi - \cos\theta\sin\phi\cos\psi & -\sin\psi\sin\phi + \cos\theta\cos\phi\cos\psi & \cos\psi\sin\theta \\ \sin\theta\sin\phi & -\sin\theta\cos\phi & \cos\theta \end{pmatrix}. \tag{8}$$

Furthermore, the angular velocity $\vec{\omega}`$ along the axes $(x', y', z')$ can be expressed in terms of the Eulerian angles and their time derivative as follows:

$$\vec{\omega}` = \begin{pmatrix} \omega_{x`} \\ \omega_{y`} \\ \omega_{z`} \end{pmatrix} = B \begin{pmatrix} \frac{d\emptyset}{dt} \\ \frac{d\theta}{dt} \\ \frac{d\psi}{dt} \end{pmatrix} , \tag{9}$$

where, according to Goldstein (1980, p. 176),

$$B = \begin{pmatrix} \sin\theta\sin\psi & \cos\psi & 0 \\ \sin\theta\cos\psi & -\sin\psi & 0 \\ \cos\theta & 0 & 1 \end{pmatrix}. \tag{10}$$

Hence from Eq.(6) we obtain

$$\vec{l} = A^{-1}\, I'\, B \begin{pmatrix} \frac{d\phi}{dt} \\ \frac{d\theta}{dt} \\ \frac{d\psi}{dt} \end{pmatrix} = A^{T}\, I'\, B \begin{pmatrix} \frac{d\phi}{dt} \\ \frac{d\theta}{dt} \\ \frac{d\psi}{dt} \end{pmatrix} \tag{11}$$

where we use $A^{-1} = A^{T} \equiv A$ transpose. Thus, we have

$$\begin{pmatrix} d\varphi/dt \\ d\theta/dt \\ d\psi/dt \end{pmatrix} = B^{-1} I'^{-1} Al = B^{-1} I'^{-1} A \begin{pmatrix} l_x \\ l_y \\ l_z \end{pmatrix}, \tag{12}$$

where

$$B^{-1} = \frac{1}{\sin\theta} \begin{pmatrix} \sin\psi & \cos\psi & 0 \\ \sin\theta\ \cos\psi & -\sin\theta\ \sin\psi & 0 \\ -\cos\theta \sin\psi & -\cos\theta\ \cos\psi & 1 \end{pmatrix}. \tag{13}$$

On the other hand, from Eq. (3) we have that

$$\vec{l} = \overrightarrow{L_0} - \mu\vec{r}\, x\, \frac{d\vec{r}}{dt}\ , \tag{14}$$

so that

$$\begin{aligned} l_x &= L_{0x} - \mu\left[ y\frac{dz}{dt} - z\frac{dy}{dt} \right], \\ l_y &= L_{0y} - \mu\left[ z\frac{dx}{dt} - x\frac{dz}{dt} \right], \\ l_z &= L_{0z} - \mu\left[ x\frac{dy}{dt} - y\frac{dx}{dt} \right]. \end{aligned} \tag{15}$$

Consequently, using Eq. (15) in Eq. (12) we obtain three additional differential equations in the unknowns $x, y, z, \phi, \theta, \psi$ and their time derivatives, and therefore together with the three equations in (2) can be integrated to provide the orbit of $m_1$ and the orientation of $m_2$ as a function of time.

## 3. Applications To A Prolate Spheroid

We have written a program to perform the above integrations numerically and applied it to the case where $m_2$ is an homogenous prolate spheroid. The expression for the force for a homogeneous prolate is calculated using the transformation

$$\begin{pmatrix} F_x \\ F_y \\ F_z \end{pmatrix} = A^{-1} \begin{pmatrix} F_{x'} \\ F_{y'} \\ F_{z'} \end{pmatrix}, \tag{16}$$

where $F_{x'}, F_{y'}, F_{z'}$ are the components of the force along the principal axes of inertia, which are given by Moulton (1914) , but are developed here in Appendices A,B,C,D and C (see Eqs. A.23, C.11, C.12, C.13 and C.15), where the more general case for an ellipsoid is also discussed.

An example of a trajectory is presented in Figure (2), where we show the evolution of a small object orbit around a prolate body during a time of days. The bodies are homogeneous with a mass density of 2300 kg/m$^3$. The shape of $m_2$, is a prolate, its semi major axis, $c$, is equal to 52 km and the eccentricity is 0.75. The initial rotational periods are given in Table 1, which are typical values for small asteroids. The smaller body is a sphere with a radius of 0.25 km . The magnitude of the initial speed and position of $m_1$ is 10 *m/s* and 9 $c$ respectively. Table 1 also shows the initial values of the Eulerian angles of the main body.

| Initial values of Euler angle a their periods | |
|---|---|
| Angle (rad) | Rotational Period |
| $\phi = 0$ | 1 hours |
| $\theta = \pi/2$ | 4 hours |
| $\psi = 0$ | 0 |

Table 1

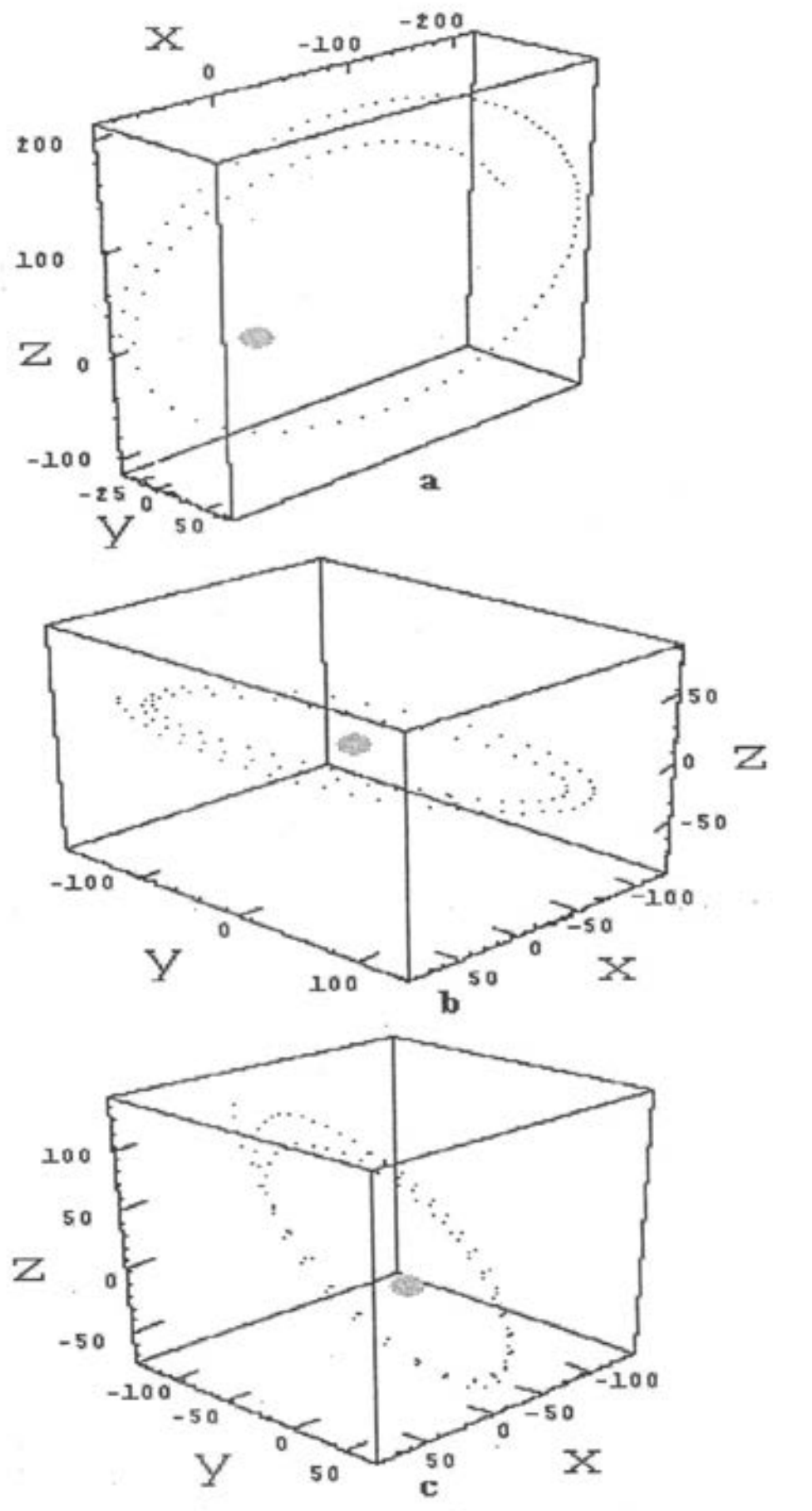

Figure 2

The three instantaneous orbits shown in Figure (2) have been constructed by iteration of 5000 position points. Each orbit in Figure (2) consists of 100 points separated by a time interval of 1000 seconds each. Therefore, the total time interval for Figure (2a) is from zero to 1.157 days. Figure (2b) has been constructed from point 2000 to 2100, or equivalent to *t*=23.140 days to *t*=24.297 days. Figure (2c) is constructed from point 4000 to 4100, or *t*=46.280 to *t*=47.437 days. This ”instantaneous” orbit shows fast changes in eccentricity and also in the position of the orbital plane. Although the intrinsic rotational effect is not shown here, for this model the main body initial rotation is affected by the orbital interaction.

## 4. Interchange Of Energy And Momentum

When $m_1$ is comparable with $m_2$, the rotation of $m_2$ is strongly affected by the proximity of $m_1$. This is illustrated in Figures $(a, b, c)$ and in the orbital video representing this solution at http://www.uprh.edu/fisica/orbita.html.

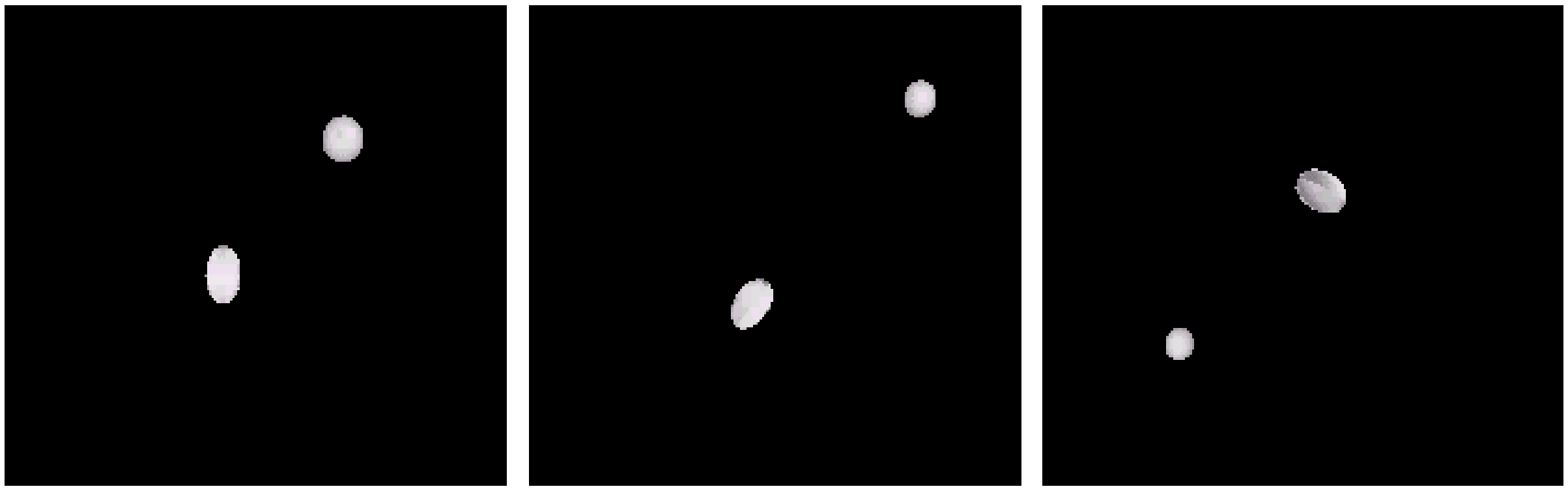

*Figures* $3_a, 3_b, 3_c$: *Satellite of Radius* $20km$, *Orbiting a Prolate of* $52km$ *Semi_major axis*

What is happening here is that the mutual torques between $m_1$ and $m_2$ changes the rotational energy $(1/2)I\omega^2$ and the intrinsic angular momentum of the prolate, $\vec{l}$. Nevertheless, the total angular momentum, $\vec{L}$, and total energy $E$, where

$$E = \frac{1}{2}\mu v^2 + U + \frac{1}{2}I\omega^2, \tag{17}$$

$$v^2 = \frac{d\vec{r}}{dt}\cdot\frac{d\vec{r}}{dt}, \tag{18}$$

remain constant, and $U$ is the potential energy of the system given by (Appendix $D$)

$$U = -m_1\pi G\rho abc\left\{\int_\lambda^\infty \frac{dv}{\Delta} - x'^2[a] - y'^2[b] - z'^2[c]\right\} =$$

$$-m_1\pi G\rho a^2 c\left\{ \frac{1}{(c^2-a^2)^{1/2}}\ln\left|\frac{1+\sqrt{\frac{c^2-a^2}{c^2+\lambda}}}{1-\sqrt{\frac{c^2-a^2}{c^2+\lambda}}}\right|\right.$$

$$-\left(x'^2 + y'^2\right)\left(\frac{\sqrt{c^2+\lambda}}{(c^2-a^2)(\lambda+a^2)} - \left(\frac{1}{2(c^2-a^2)^{3/2}}\right)\ln\left|\frac{1+\sqrt{\frac{c^2-a^2}{c^2+\lambda}}}{1-\sqrt{\frac{c^2-a^2}{c^2+\lambda}}}\right|\right)$$

$$\left.-z'^2\left(\left(\frac{1}{(c^2-a^2)^{3/2}}\right)\ln\left|\frac{1+\sqrt{\frac{c^2-a^2}{c^2+\lambda}}}{1-\sqrt{\frac{c^2-a^2}{c^2+\lambda}}}\right| - 2\left(\frac{1}{(c^2-a^2)\sqrt{c^2+\lambda}}\right)\right)\right\}, \tag{19}$$

where

$$\lambda = \frac{x'^2+y'^2+z'^2-(c^2-a^2)-2a^2+\sqrt{\{x'^2+y'^2+z'^2-(c^2-a^2)\}^2+4(c^2-a^2)(x'^2+y'^2)}}{2} \tag{20}$$

$$a \equiv semi_minor\ axis \tag{21}$$

$$c \equiv semi_mayor\ axis \tag{22}$$

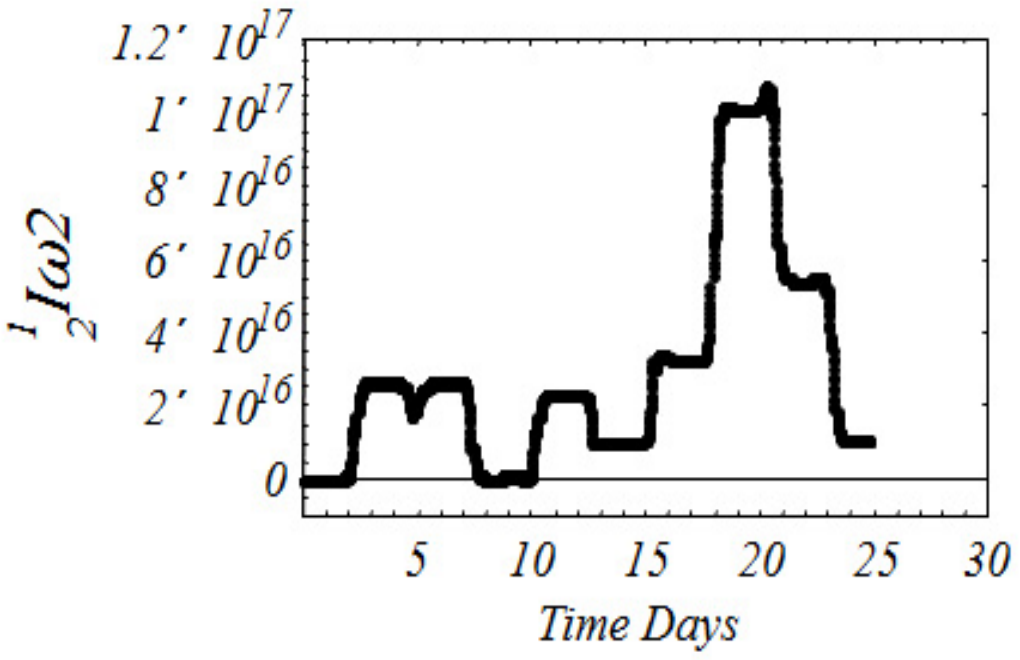

Figure 4a. Time evolution of the intrinsic rotational energy for the system of comparable masses.

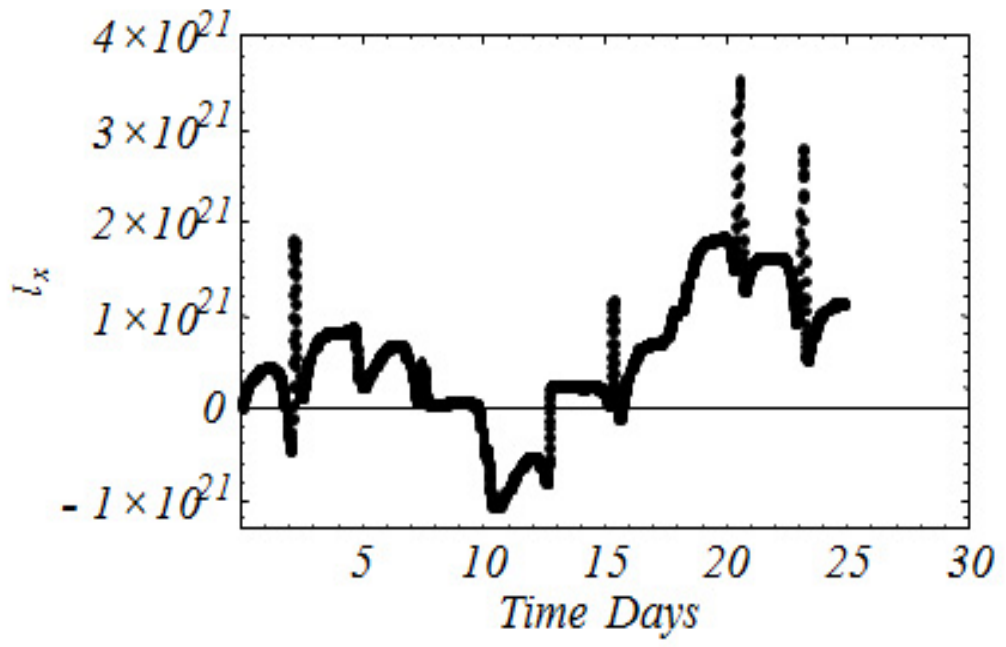

Figure 4b. Time evolution of the intrinsic angular momentum component $l_x$, for the system of comparable masses.

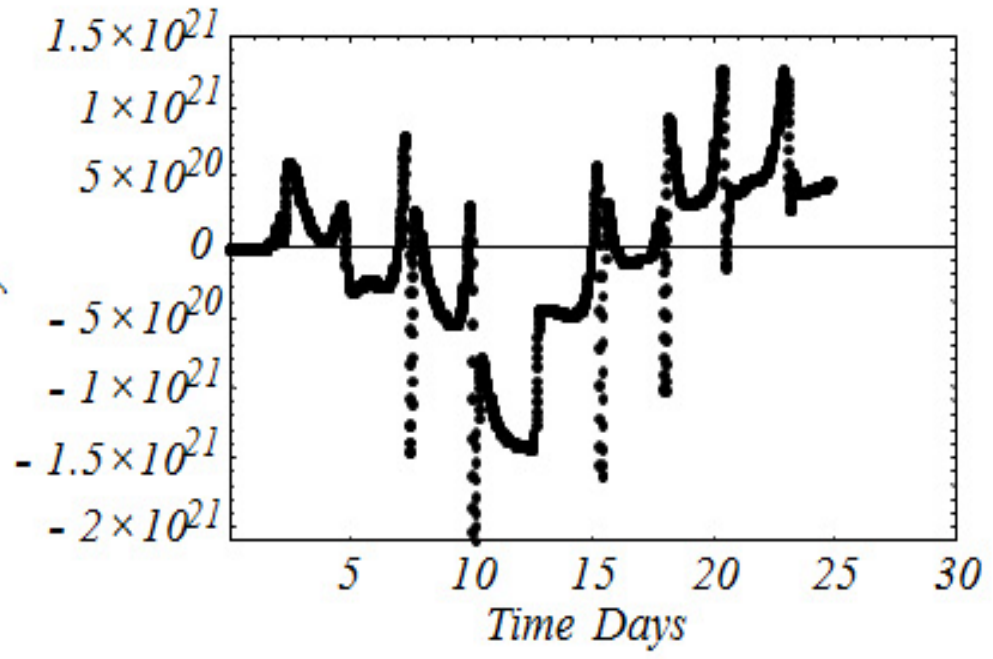

Figure 4c. Time evolution of the intrinsic angular momentum component $l_y$, for the system of comparable masses.

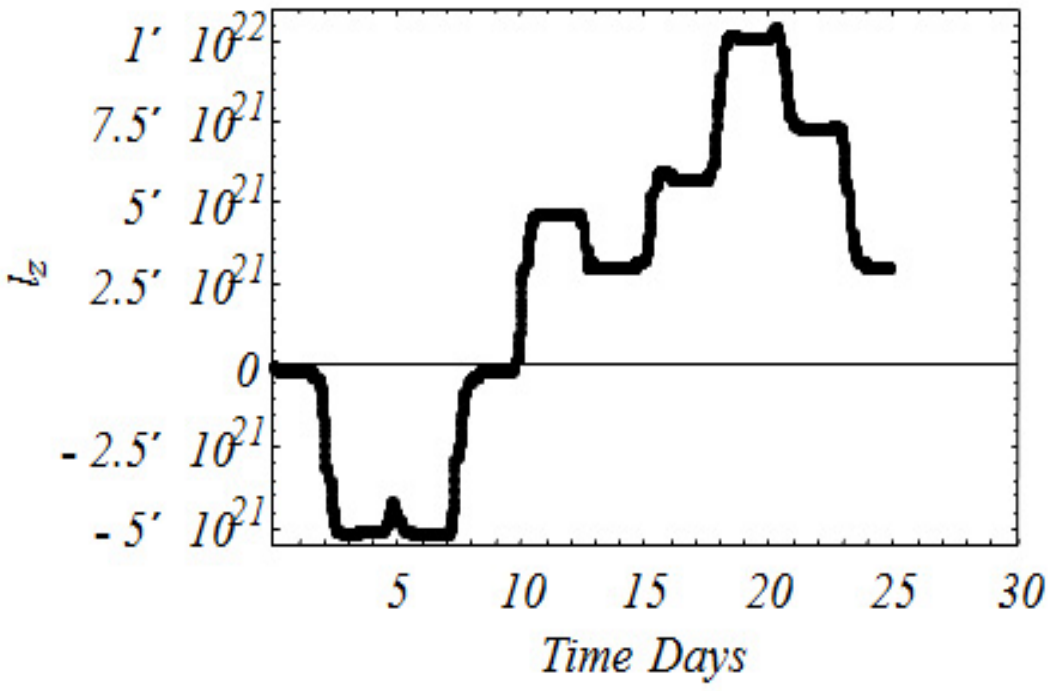

Figure 4d. Time evolution of the intrinsic angular momentum component $l_z$, for the system of comparable masses.

In Figure (4) we can see the time behavior of the intrinsic angular energy, $(1/2)I\omega^2$, and the intrinsic angular momentum components, $l_x$, $l_y$, $l_z$, for the system shown in Figure (3). In Figure (5), the conservation of $E$ and $L_x$, $L_y$, $L_z$ is also illustrated. It is worth noticing that the explicit verification of the conservation of $E$ and $L$ (Figures (5) and (7)) is a direct check on the correctness of the theoretical analysis and computations.

On the other hand, the total orbital energy of the system

$$E_{orb} = \frac{1}{2}\mu v^2 + U, \tag{22}$$

changes with time and could become positive, and thus an escape velocity could be given to a satellite. Alternatively, for sufficiently negative values of $E_{orb}$, collisions between $m_1$ and $m_2$ will occur. Figure (6a) shows the behavior of the $E_{orb}$ vs. time for a system consisting of a prolate with a semi-major axis and eccentricity of 52 *km* and 0.75 respectively. The spherical satellite has a radio of 7 *km*, initial speed of 12 *m/s* and a position of 4 $c$; both bodies have the same mass density of $2300 kg/m^3$. The zigzag steps in the functional behavior of the curve $E_{orb}$ vs. time are a consequence of the fact that the interchanges between orbital energy and angular energy occur mostly during the relatively short time that the satellite spends near to the prolate. We also see that eventually, after about 105 days, $E_{Orb}$ becomes positive and constant, and the satellite continuously separates from the prolate. Figure (8) presents the explicit trajectory of this escaping satellite. As a matter of fact, Figures (6b) and (6c) show that after about 105 days, $\frac{1}{2} I\omega^2$ and $\vec{l}$ become constants and the satellite decouples from the prolate dynamics. Further discussions of this mechanism for transferring angular energy into orbital energy are addressed in the final section.

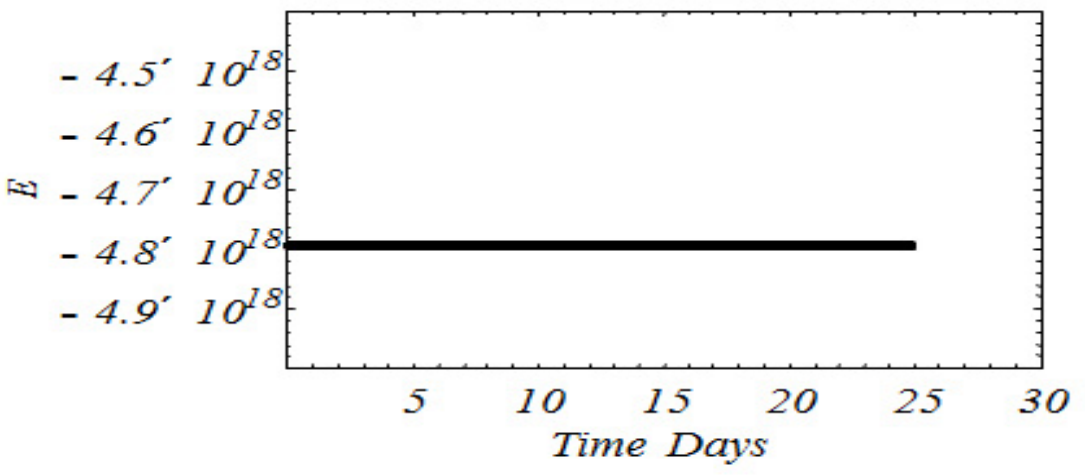

Figure 5a. The value of the total energy , $E$, for the system of comparable masses.

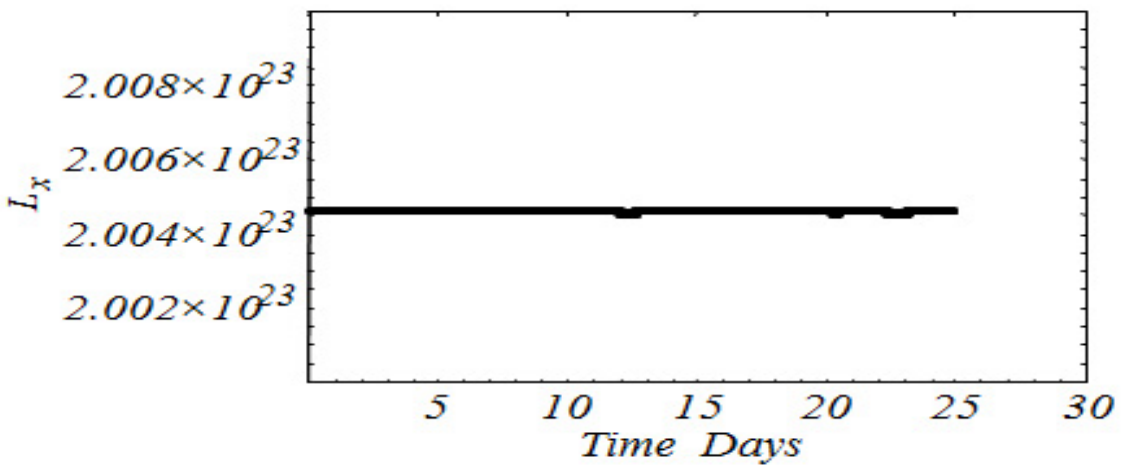

Figure 5b. The value for total momentum component $L_y$, for the system of comparable

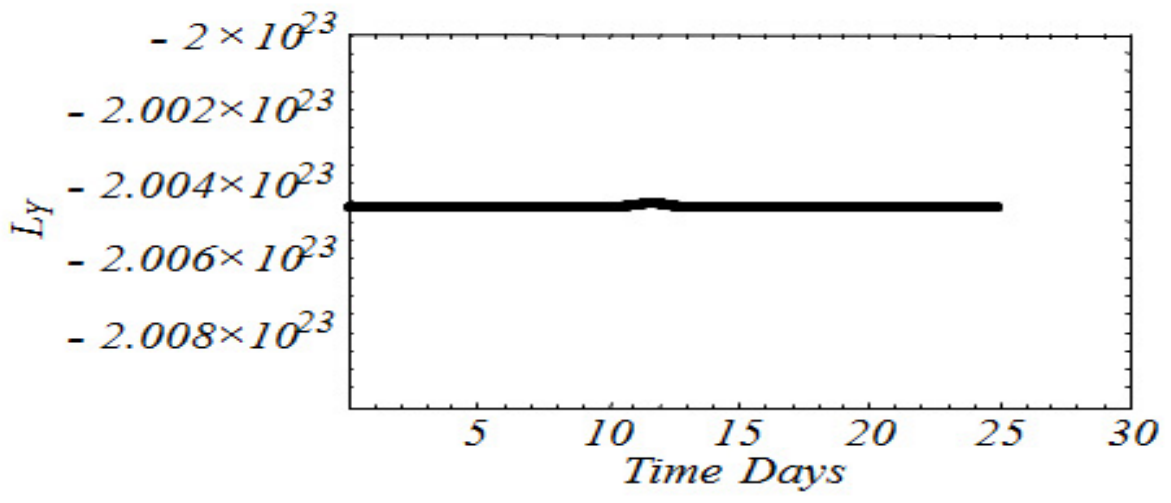

Figure 5c. The value for total momentum component $L_y$, for the system of comparable masses

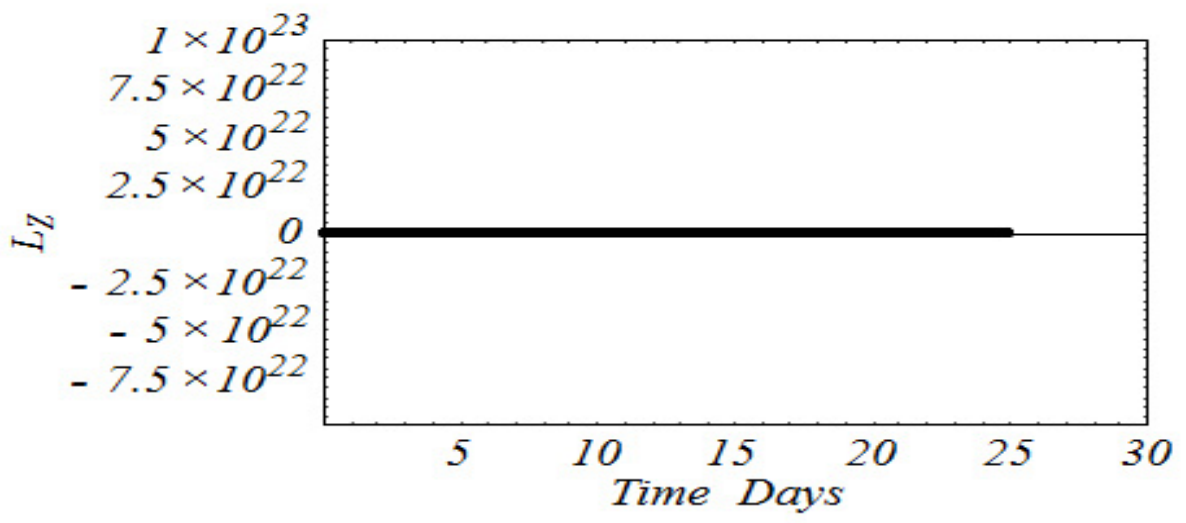

Figure 5d. The value for total momentum component $L_z$, for the system of comparable masses

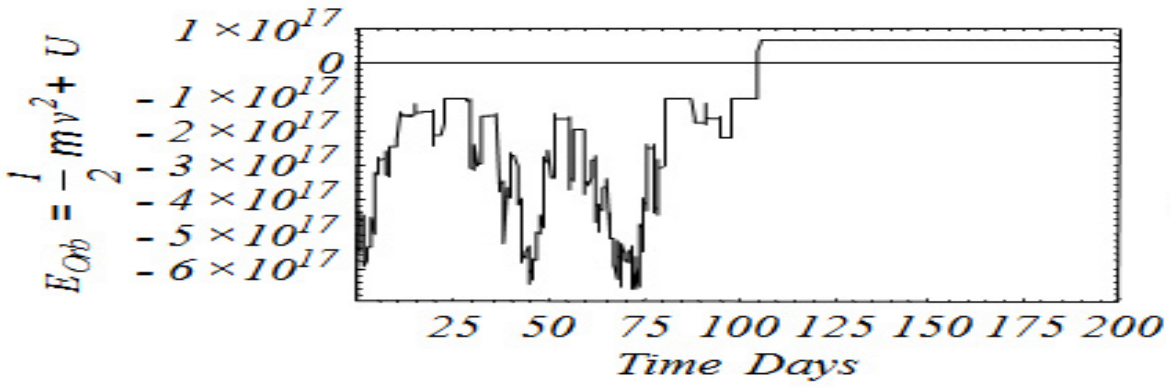

Figure 6a. Time evolution of the orbital energy $E_{orb}$, for escaping satellite.

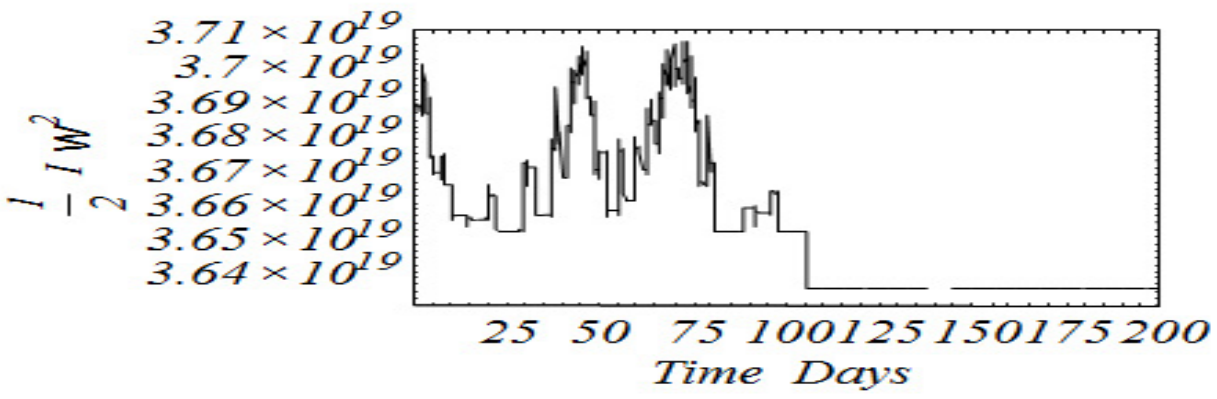

Figure 6b. Time evolution of the intrinsic rotational energy of the proloid for the system with an escaping satellite.

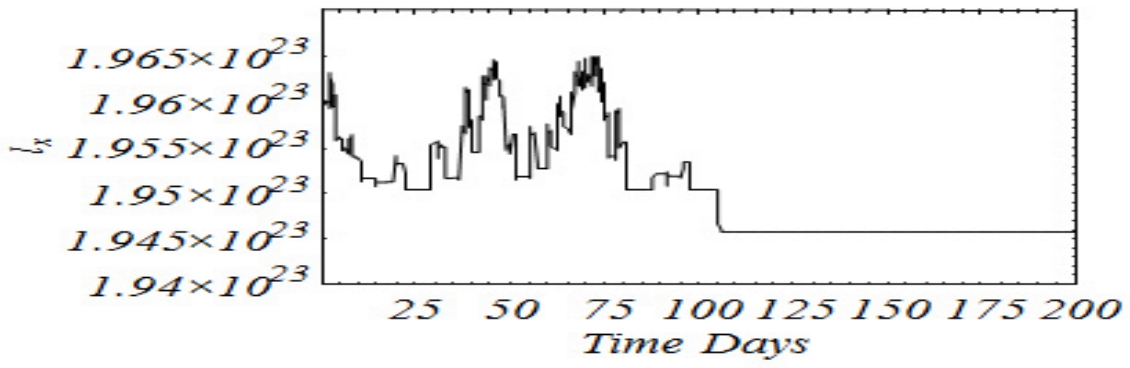

Figure 6c. Time evolution of the intrinsic angular momentum $l_x$, for the system with an escaping satellite.

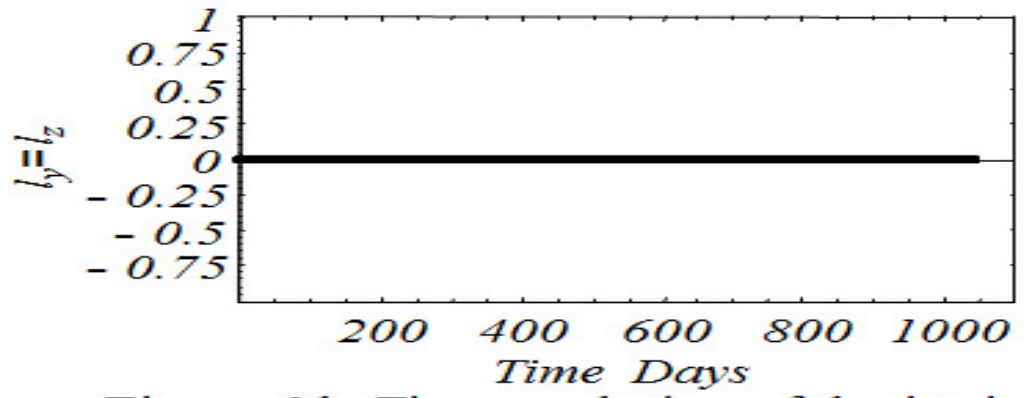

Figure 6d. Time evolution of the intrinsic angular momentum components $l_y$ and $l_z$, for the system with an escaping satellite

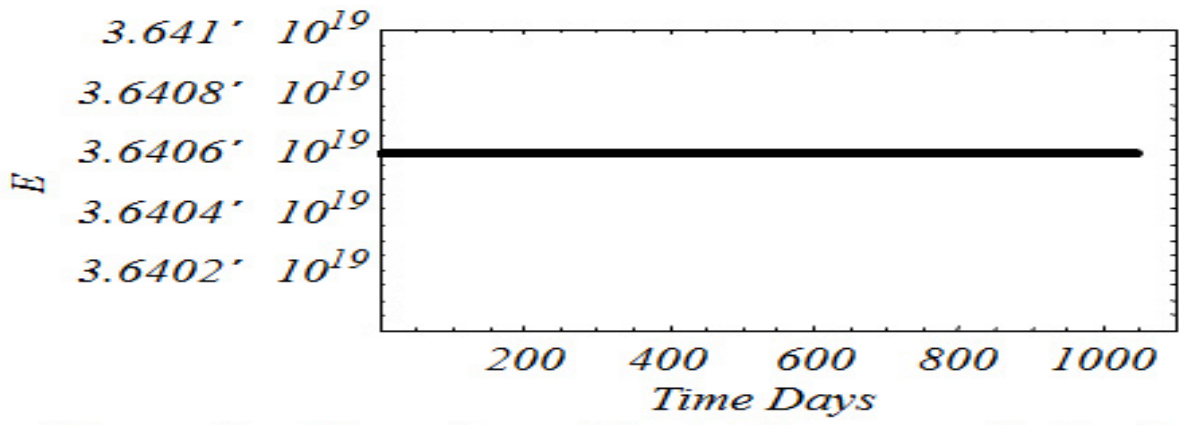

Figure 7a. The value of the total energy, $E$, for the system with an escaping satellite.

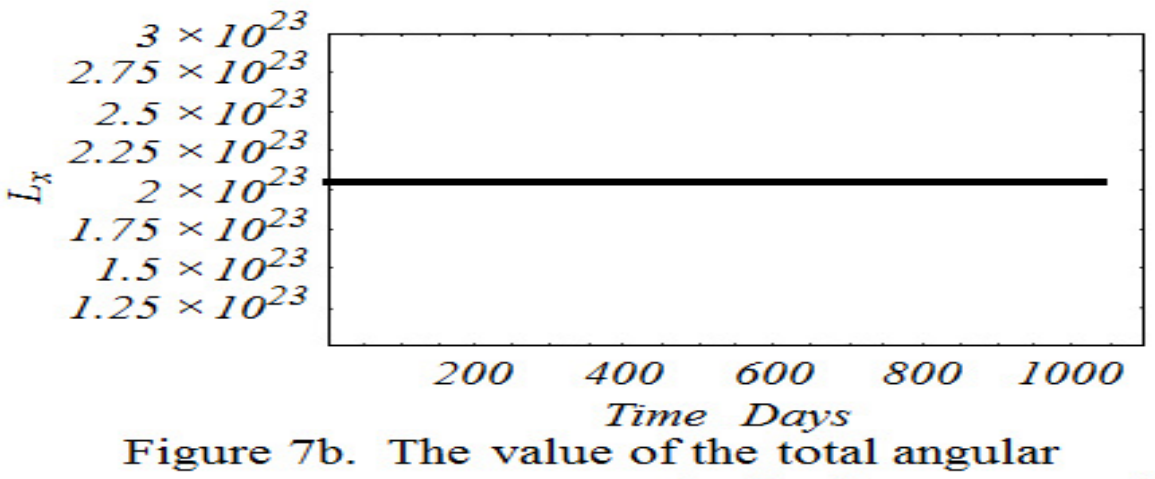

Figure 7b. The value of the total angular momentum component $L_x$, for the system with escaping satellite.

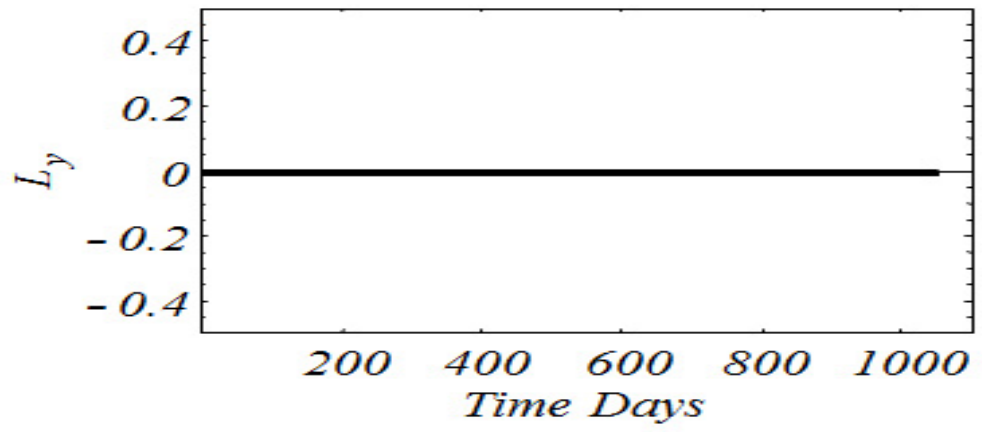

Figure 7c. The value of the total angular momentum component $L_y$, for the system with escaping satellite.

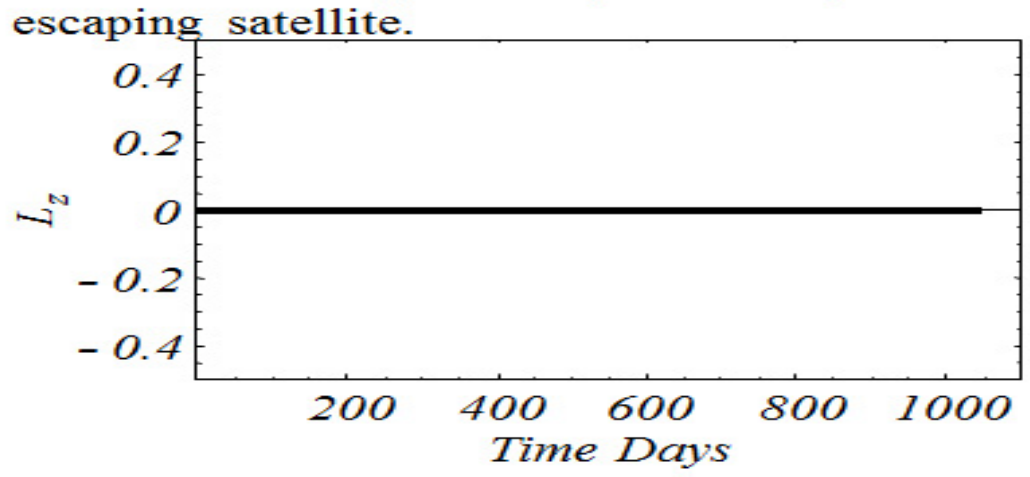

Figure 7d. The value of the total angular momentum component $L_z$, for the system with escaping satellite.

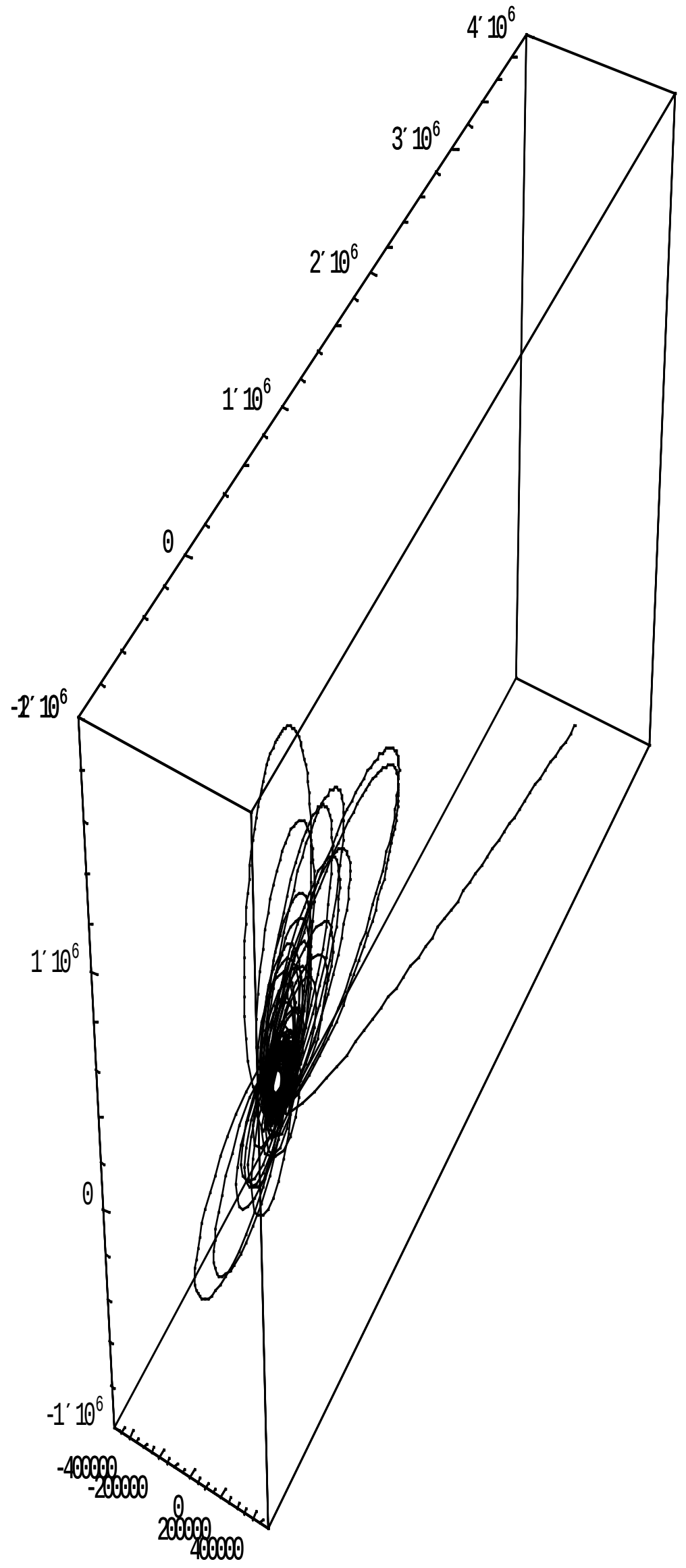

Figure 8. Orbit of the escaping satellite. Distance in meters.

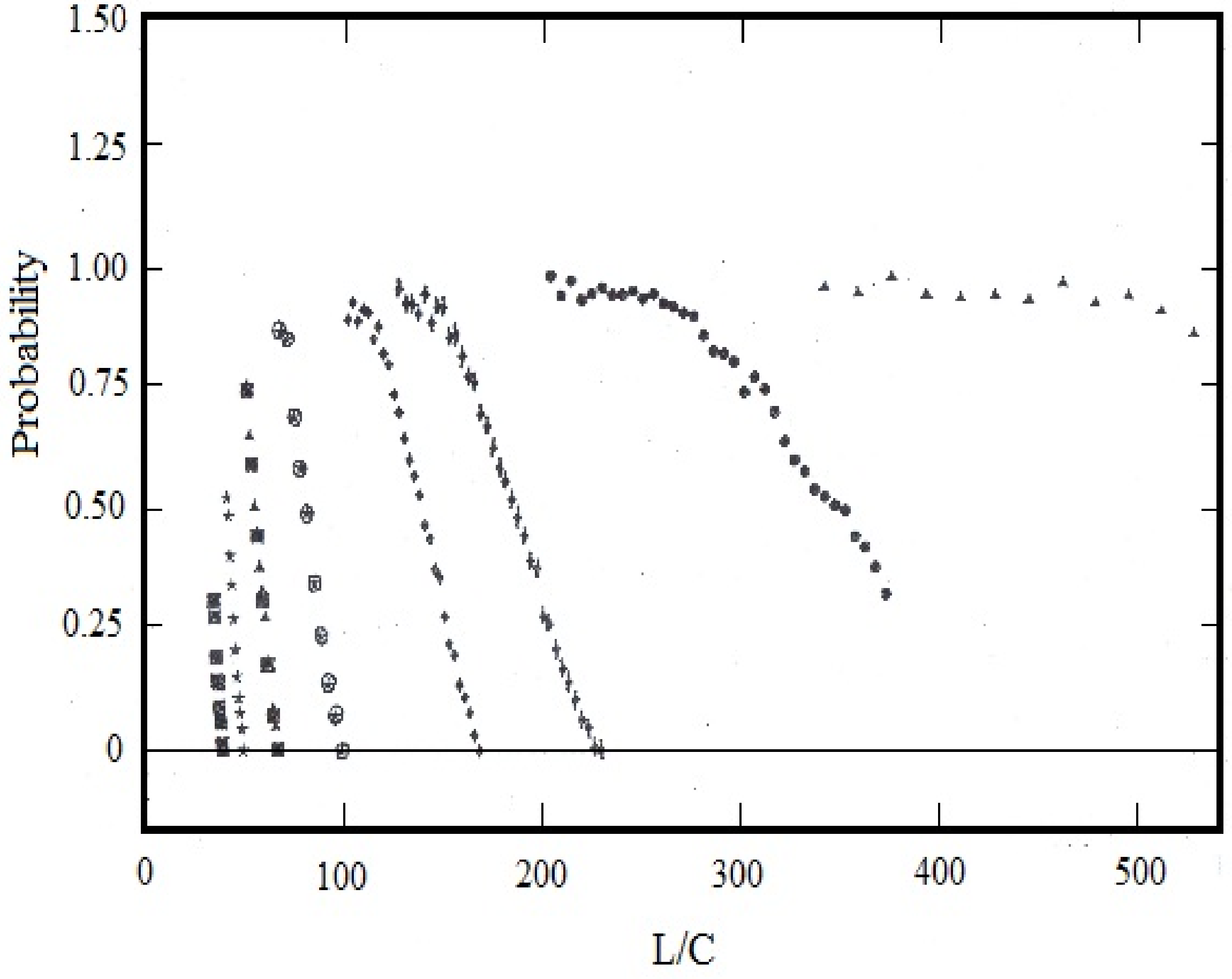

Figure 9

Probability for the capture of a spherical body ,$P$ ,of radius 50 $km$ , by an initially non rotating prolate of semi-mayor axis $c = 50\,km$ . Each set of symbols ,from left to right, represents the following relative initial speeds: $v = \frac{3.0m}{s}, \frac{2.5m}{s}, \frac{2.0m}{s}, \frac{1.5m}{s}, \frac{1.0m}{s}, \frac{0.8m}{s}, \frac{0.5m}{s}, \frac{0.3m}{s}$. For instance, the squares at the extreme left of Figure 9 represent the probability as a function of the parameter of impact $L$ , in units of semi- mayor axis, $c$, for $v = \frac{3.0m}{s}$, while the last set with triangles, at the extreme right, has $v = \frac{0.3m}{s}$.

The almost straight line part of the curves in Figure (9), that starts at $P = 0$, can be described by the equation

$$P = 3.5 - x \tag{23}$$

where

$$x = \left(\frac{2.1}{100}\right) v^{\frac{4}{3}} \left(\frac{L}{c}\right) \quad , \tag{24}$$

and therefore the probability for capture for all curves in this region is given by the same linear function, namely Eq.(23), of $x$ . We also find that the abrupt end of each curve, before reaching a probability 1, is due to the physical collision of the binary system, and is approximately satisfied by the equation

$$\frac{L}{c} = 100/v \quad . \tag{25}$$

We are grateful to Dr. Samuel Vazquez , at the Perimeter Institute for Theoretical Physics, Ontario, Canada, for the Monte Carlo simulation of Figure (9) and his very helpful suggestions.

## 5. Escaping Satellites

The velocity of the satellite relative to an origin at the Sun center is equal to the velocity of the satellite relative to the binary system center of mass, $\vec{V}_{Sat}$, plus the velocity of this center of mass relative to the Sun center $\vec{V}_0$. Hence, the decoupled satellite will have an orbit around the Sun with a different eccentricity and semi-major axis, and therefore the new perihelion could be larger or smaller than that in the original binary system.

For example, the perihelion, $r_p$, for the new orbit when $\vec{V}_0$ and $\vec{V}_{Sat}$ are in opposite directions and the satellite is escaping at the aphelion, $r_a$, is given by (Appendix F)

$$r_p = r_a \left[ \frac{(1-\in)\left(1-\frac{V_{Sat}}{V_0}\right)^2}{2-(1-\in)\left(1-\frac{V_{Sat}}{V_0}\right)^2} \right]; \qquad V_{Sat} = \left|\vec{V}_{Sat}\right|, \; V_0 = \left|\vec{V}_0\right|, \tag{26}$$

where $\in$ is the eccentricity of the binary system center of mass orbit, and $V_0$ at aphelion can be calculated ( see Appendix F ,Eq. F.7) from

$$V_0 = \sqrt{\frac{GM_{\circ}(1-\in)}{r}}. \tag{27}$$

For instance, an original binary orbit in the asteroid belt with $r_a = 4A.U.$ and $\in = 1/3$ has $V_0 = (1.21)10^4 \frac{m}{s}$ . On the other hand, we have calculated, for the example of Figure (6), that the satellite escape after 105 days with $V_{sat} = 5.70\, m/s$, and consequently $V_{sat}/V_0 \approx 0$ , thus Eq. (26) gives essentially the same perihelion as the original orbit, namely $\frac{r_a(1-\in)}{1+\in}=2A.U.$ However, situations where $V_{Sat}\cong V_0$ would mean that $\frac{r_p}{r_a}\cong 0$, and that could happen to binary comets in the Oort cloud. For example, if $r=10^5 A.U.$, we find from Eq. (27) that

$$V_0=94\frac{m}{s}(1-\in), \quad (28)$$

and thus the condition $V_0\approx V_{Sat}\approx 5.70\frac{m}{s}$ would imply that $\in\cong 0.9$. Hence, it is plausible that satellites of comets with very high eccentricity in the Oort cloud could be sent into Earth crossing trajectories by this mechanism.

## Appendix A: The Field And The Potential Of A Homogeneous Ellipsoid

It can be shown that the potential exterior to a constant density, $\rho$, ellipsoid can be written in the form ( J. M. A. Dandy ,p 110):

$$V=-\pi G\rho abc\{\int_\lambda^\infty \frac{dv}{\Delta}-x^{`2}[a]-y^{`2}[b]-z^{`2}[c]\} \quad , \qquad \text{A.1}$$

where

$$\Delta\equiv\sqrt{(a^2+v)(b^2+v)(c^2+v)} \quad , \qquad \text{A.2}$$

$$[a_i]\ \equiv\ \int_\lambda^\infty \frac{dv}{(a_i{}^2+v)\Delta}, \quad a_1=a\,,\ a_2=b\,,\ a_3=c \quad , \qquad \text{A.3}$$

and λ is given by the real positive solution of the following cubic equation:

$$\frac{x^{`2}}{\lambda+a^2}+\frac{y^{`2}}{\lambda+b^2}+\frac{z^{`2}}{\lambda+c^2}=1 \quad . \qquad \text{A.4}$$

Note that $\lambda=0$ corresponds to points at the surface of the ellipsoid , with semi-principal axes $a,b,c$. We have determined, in Appendix D, that the solution for λ is

$$\lambda=2\left(\sqrt{\frac{\bar{a}}{3}}\right)T\left[1/3,\frac{\bar{b}}{2\sqrt{(\frac{\bar{a}}{3})^3}}\right]-p/3 \quad , \qquad \text{A.5}$$

where $T[n,x]$ is the Chebyshev functions of the first kind (Mathematica notation),which satisfy

$$T[n,cos\theta]=\cos n\theta\ , \qquad \text{A.6}$$

and

$$\bar{a}=\left(\frac{1}{3}\right)(p^2-3q)\ , \qquad \text{A.7}$$

$$\bar{b} = \left(\frac{1}{27}\right)(9pq - 2p^3 - 27r) \quad , \qquad \text{A.8}$$

$$p = a^2 + b^2 + c^2 - x`^2 - y`^2 - z`^2 \quad , \qquad \text{A.9}$$

$$q = a^2b^2 + a^2c^2 + b^2c^2 - (c^2 + b^2)x`^2 - (c^2 + a^2)y`^2 - (a^2 + b^2)z`^2 \ , \qquad \text{A.10}$$

$$r1 = a^2b^2c^2 - (c^2b^2)x`^2 - (c^2a^2)y`^2 - (a^2b^2)\, z`^2 \quad . \qquad \text{A.11}$$

The components of the gravitational field , $\vec{g}$ , in the principal axes are given by:

$$g_{x`_i} = -\frac{\partial V}{\partial x`_i} = -2\pi G\rho abcx`_i \ [a_i] \quad , \ x`_1 = x` \, , \ x`_2 = y` \, , \ x`_3 = z` \ . \qquad \text{A.12}$$

The above relationship follows from differentiating , with respect to $x`_i$ , Eq. A.1 and then using Eq. A.4, together with the mathematical fact that

$$\frac{\partial}{\partial x`_i} \int_\lambda^\infty f(v)dv = -\frac{\partial \lambda}{\partial x`_i} f(\lambda) \quad . \qquad \text{A.13}$$

Note that $V$ can be rewritten in the interesting form

$$V = -\pi G\rho abc \int_\lambda^\infty \frac{dv}{\Delta} - (\tfrac{1}{2})\vec{r}` . \vec{g} \quad , \qquad \text{A.14}$$

where $\vec{r}` = (x`, y`, z`)$ is the position vector.

The potential and the field can be rewritten in terms of the elliptic integrals of the first kind, $F$, the second kind, $E$ , and the third kind, $\Pi$ , (Appendix B) by using(Mathematica notation) :

$$[a] = \frac{2}{(c^2-a^2)^{3/2}} \left[\Pi(1; \alpha|s) - F(\alpha|s)\right] \quad , \qquad \text{A.15}$$

$$[b] = \frac{2}{s(c^2-a^2)^{3/2}} \left[\Pi(s; \alpha|s) - F(\alpha|s)\right] \quad , \qquad \text{A.16}$$

$$[c] = \frac{2}{s(c^2-a^2)^{3/2}} \left[F(\alpha|s) - E(\alpha|s)\right] \quad , \qquad \text{A.17}$$

$$\int_\lambda^\infty \frac{dv}{\Delta} = \frac{2}{(c^2-a^2)^{1/2}} F(\alpha|s) \quad , \qquad \text{A.18}$$

where

$$\alpha = \sin^{-1}\sqrt{\frac{c^2-a^2}{c^2+\lambda}} \quad , \qquad \text{A.19}$$

$$s = \frac{c^2-b^2}{c^2-a^2} \quad , \qquad \text{A.20}$$

$$a \leq b < c \quad . \qquad \text{A.21}$$

We then have that for a test particle or a spherical mass,  , the potential energy , $U$ , and the force , $\vec{F}`$ , are then given by

$$U = mV \quad , \qquad \text{A.22}$$

$$\vec{F}` = m\vec{g} \quad , \qquad \text{A.23}$$

and the force components in a frame $x, y, z,$ with axes fixed in space are then obtained using Eq. (16).

## Appendix B: Expressing The Potential And The Field In Terms Of Elliptic Integrals

The elliptic integrals of the first kind, $F$, the second kind, $E$, and the third kind, $\Pi$, are defined as (Mathematica notation) :

$$F(\alpha|m) = \int_0^\alpha [1 - msin^2\theta]^{-1/2}\, d\theta \quad , \qquad \text{B.1}$$

$$E(\alpha|m) = \int_0^\alpha [1 - msin^2\theta]^{1/2}\, d\theta \quad , \qquad \text{B.2}$$

$$\Pi(n;\alpha|m) = \int_0^\alpha \, [1 - nsin^2\theta]^{-1}\, [1 - msin^2\theta]^{-1/2}\, d\theta \quad . \qquad \text{B.3}$$

Furthermore , we have from Eqs. B.1 and B.2 .

$$\Pi - F = \int_0^\alpha nsin^2\theta\, [1 - nsin^2\theta]^{-1}\, [1 - msin^2\theta]^{-1/2}\, d\theta \quad . \qquad \text{B.4}$$

With the substitution

$$\sin\theta = (\gamma v + \beta)^{-1/2}\, , \gamma, \beta, constants \quad , \qquad \text{B.5}$$

Eq. B.4 becomes

$$\Pi - F = \frac{n}{2\gamma^{3/2}} \int_{v_0}^{\infty} \frac{dv}{\left(\frac{\beta-n}{\gamma}+v\right)(\frac{\beta}{\gamma}+v)^{1/2}\,(\frac{\beta-1}{\gamma}+v)^{1/2}\,(\frac{\beta-m}{\gamma}+v)^{1/2}} \quad , \qquad \text{B.6}$$

where

$$v_{\,0} = \frac{sin^{-2}\alpha - \beta}{\gamma} \quad , \qquad \text{B.7}$$

or

$$sin\alpha = (\gamma v_0 + \beta)^{-1/2} \quad . \qquad \text{B.8}$$

Now, let

$$a^2 \equiv \frac{\beta-1}{\gamma} \quad , \qquad \text{B.9}$$

$$b^2 \equiv \frac{\beta-m}{\gamma} \quad , \qquad \text{B.10}$$

$$c^2 \equiv \frac{\beta}{\gamma} \quad , \qquad \text{B.11}$$

$$l^2 \equiv \frac{\beta - n}{\gamma} \quad , \qquad \text{B.12}$$

$$\lambda \equiv v_0 \quad , \qquad \text{B.13}$$

then we find , using B.9, B.10 and B.11, that

$$\gamma = \frac{1}{c^2 - a^2} \quad , \qquad \text{B.14}$$

$$m = \frac{c^2 - b^2}{c^2 - a^2} \equiv s \quad . \qquad \text{B.15}$$

After substitution of Eqs. B.9 to B.14 into Eq. B.6, we obtain

$$\Pi - F = \frac{n(c^2 - a^2)^{3/2}}{2} \int_\lambda^\infty \frac{dv}{(l^2+v)\,(c^2+v)^{1/2}\,(\,a^2+v)^{1/2}\,(\,b^2+v)^{1/2}} \quad , \qquad \text{B.16}$$

or, using the definition A.2 and

$$[l\,] \ \equiv \ \int_\lambda^\infty \frac{dv}{(l^2+v)\Delta}, \qquad \text{B.17}$$

we can rewrite Eq. B.16 as

$$\Pi - F = \ \frac{n(c^2 - a^2)^{3/2}}{2} [l\,] \quad , \qquad \text{B.18}$$

from which

$$[l\,] = \frac{2(\Pi - F)}{n(c^2 - a^2)^{3/2}} \quad . \qquad \text{B.19}$$

If $n = 1$, we see, from Eqs. B.12 and B.9, that $l^2 = a^2$ and Eq. B.19 becomes Eq. A.15. On the other hand, if $n = m = s$, we have, from Eqs. B.10 and B.12, that $l^2 = b^2$ and therefore Eq. B.19 reproduces Eq. A.16.

In order to obtain Eq. A.17 consider Eqs. B.1 and B.2 which imply that

$$F - E = \int_0^\alpha m sin^2\theta \ \ [1 - m sin^2\theta]^{-1/2}\, d\theta \quad . \qquad \text{B.20}$$

Moreover, we see from Eqs. B.4 and B.15 that

$$\lim_{n\to 0}(\Pi - F)/n) \to \qquad \int_0^\alpha sin^2\theta \ \ [1 - m sin^2\theta]^{-1/2}\, d\theta = (F - E)/s \quad , \qquad \text{B.21}$$

where we use $m = s$. However, in the limit $n \to 0$, we have, using Eqs. B.11 and B.12, $l^2 = c^2$, and therefore, using Eqs. B.19 and B.21,

$$\lim_{n\to 0}[l\,] = [c\,] = \frac{2(F - E)}{s(c^2 - a^2)^{3/2}} \quad , \qquad \text{B.22}$$

which is Eq. A.17.

Finally , using the transformation B.5 in Eq. B.1, we can write

$$F = \frac{1}{2\gamma^{1/2}} \int_{\lambda}^{\infty} \frac{dv}{(c^2+v)^{1/2}\,(a^2+v)^{1/2}\,(b^2+v)^{1/2}} = \frac{1}{2\gamma^{1/2}} \int_{\lambda}^{\infty} \frac{dv}{\Delta} \quad , \qquad \text{B.23}$$

or by inverting the above equation, we obtain Eq. A.18:

$$\int_{\lambda}^{\infty} \frac{dv}{\Delta} = 2\gamma^{1/2} F = \frac{2F}{(c^2-a^2)^{1/2}} \ , \qquad \text{B.24}$$

where we used Eq. B.14.

## Appendix C: The Potential And The Field For A Homogeneous Prolate

If $a = b$, we have a homogeneous prolate, and from Eq. A.20 we have $s = 1$ ,which implies , from Eqs. A.15 and A.16 , that

$$[a] = [b] = \frac{2}{(c^2-a^2)^{3/2}} \left[\Pi(1;\alpha|1) - F(\alpha|1)\right] \quad . \qquad \text{C.1}$$

Furthermore, using Eq. B.1 with $m = s = 1$, we obtain

$$F(\alpha|1) = \int_0^{\alpha} [1 - sin^2\theta]^{-1/2}\, d\theta = (1/2)\ln\left[\frac{1+sin\alpha}{1-sin\alpha}\right] \quad , \qquad \text{C.2}$$

And hence, using Eq. A.19,

$$F(\alpha|1) = (1/2)\ln\left[\frac{1+\sqrt{\frac{c^2-a^2}{c^2+\lambda}}}{1-\sqrt{\frac{c^2-a^2}{c^2+\lambda}}}\right] \quad . \qquad \text{C.3}$$

On the other hand, Eq. B.3 tells us that

$$\Pi(1;\alpha|1) = \int_0^{\alpha} [1 - sin^2\theta]^{-3/2}\, d\theta = \left(\frac{1}{2}\right)\frac{sin\alpha}{cos^2\alpha} + \left(\frac{1}{4}\right) ln\left[\frac{1+sin\alpha}{1-sin\alpha}\right] \ , \qquad \text{C.4}$$

which again after using A.19 and some algebraic manipulations becomes

$$\Pi(1;\alpha|1) = \frac{(\frac{1}{2})\sqrt{c^2-a^2}\sqrt{c^2+\lambda}}{\lambda+a^2} + (1/4)\ln\left[\frac{1+\sqrt{\frac{c^2-a^2}{c^2+\lambda}}}{1-\sqrt{\frac{c^2-a^2}{c^2+\lambda}}}\right] \quad . \qquad \text{C.5}$$

Therefore, from Eqs. C.1, C.3 and C.5, we get

$$[a] = [b] = \frac{2}{(c^2-a^2)^{3/2}}\left(\frac{(\frac{1}{2})\sqrt{c^2-a^2}\sqrt{c^2+\lambda}}{\lambda+a^2} - (1/4)\ln\left[\frac{1+\sqrt{\frac{c^2-a^2}{c^2+\lambda}}}{1-\sqrt{\frac{c^2-a^2}{c^2+\lambda}}}\right]\right) \quad . \qquad \text{C.6}$$

Finally, Eq. A.17 with $s = 1$ is

$$[c] = \frac{2}{(c^2-a^2)^{3/2}} \left[F(\alpha|1) - E(\alpha|1)\right] \quad , \qquad \text{C.7}$$

and with Eq. B.2,

$$E(\alpha|1) = \int_0^{\alpha}[1 - sin^2\theta]^{1/2}\, d\theta = sin\alpha = \sqrt{\frac{c^2-a^2}{c^2+\lambda}} \quad . \qquad \text{C.8}$$

Thus, we have

$$[c] = \frac{2}{(c^2-a^2)^{3/2}}\left(\left(\frac{1}{2}\right)\ln\left[\frac{1+\sqrt{\frac{c^2-a^2}{c^2+\lambda}}}{1-\sqrt{\frac{c^2-a^2}{c^2+\lambda}}-}\right] - \sqrt{\frac{c^2-a^2}{c^2+\lambda}}\right), \qquad \text{C.9}$$

and from Eq. A.18, using Eq. C.3 ,

$$\int_\lambda^\infty \frac{dv}{\Delta} = \frac{2}{(c^2-a^2)^{1/2}}\, F(\alpha|1) = \frac{1}{(c^2-a^2)^{1/2}}\ln\left[\frac{1+\sqrt{\frac{c^2-a^2}{c^2+\lambda}}}{1-\sqrt{\frac{c^2-a^2}{c^2+\lambda}}-}\right] . \qquad \text{C.10}$$

Then the substitution of Eqs. C.6, C.9 and C.10 in Eq. A.12 gives

$$g_{x`} = -2\pi G\rho a^2 cx`[a] = -2\pi G\rho a^2 cx`\left(\frac{\sqrt{c^2+\lambda}}{(c^2-a^2)(\lambda+a^2)} - (\frac{1}{2(c^2-a^2)^{3/2}})\ln\left[\frac{1+\sqrt{\frac{c^2-a^2}{c^2+\lambda}}}{1-\sqrt{\frac{c^2-a^2}{c^2+\lambda}}-}\right]\right) , \qquad \text{C11}$$

$$g_{y`} = -2\pi G\rho a^2 cy`[b] = -2\pi G\rho a^2 cy`\left(\frac{\sqrt{c^2+\lambda}}{(c^2-a^2)(\lambda+a^2)} - \left(\frac{1}{2(c^2-a^2)^{\frac{3}{2}}}\right)\ln\left[\frac{1+\sqrt{\frac{c^2-a^2}{c^2+\lambda}}}{1-\sqrt{\frac{c^2-a^2}{c^2+\lambda}}-}\right]\right) , \qquad \text{C12}$$

$$g_{z`} = -2\pi G\rho a^2 cz`[c] = -2\pi G\rho a^2 cz`\left(\left(\frac{1}{(c^2-a^2)^{3/2}}\right)\ln\left[\frac{1+\sqrt{\frac{c^2-a^2}{c^2+\lambda}}}{1-\sqrt{\frac{c^2-a^2}{c^2+\lambda}}-}\right] - \frac{2}{(c^2-a^2)}\sqrt{\frac{1}{(c^2+\lambda)}}\right) . \qquad \text{C13}$$

Also the potential is , using Eqs. A.1, C.6, C.9 and C.10,

$$V = -\pi G\rho abc\left\{\int_\lambda^\infty \frac{dv}{\Delta} - x`^2[a] - y`^2[b] - z`^2[c]\right\} =$$

$$-\pi G\rho a^2 c\left\{ \frac{1}{(c^2-a^2)^{1/2}}\ln\left[\frac{1+\sqrt{\frac{c^2-a^2}{c^2+\lambda}}}{1-\sqrt{\frac{c^2-a^2}{c^2+\lambda}}-}\right]\right.$$

$$- (x`^2 + y`^2)\left(\frac{\sqrt{c^2+\lambda}}{(c^2-a^2)(\lambda+a^2)} - (\frac{1}{2(c^2-a^2)^{3/2}})(\ln\left[\frac{1+\sqrt{\frac{c^2-a^2}{c^2+\lambda}}}{1-\sqrt{\frac{c^2-a^2}{c^2+\lambda}}-}\right]\right)$$

$$\left.-z`^2\left(\left(\frac{1}{(c^2-a^2)^{3/2}}\right)\ln\left[\frac{1+\sqrt{\frac{c^2-a^2}{c^2+\lambda}}}{1-\sqrt{\frac{c^2-a^2}{c^2+\lambda}}-}\right] - 2(\frac{1}{(c^2-a^2)\sqrt{c^2+\lambda)}})\right)\right\} . \qquad \text{C.14}$$

Finally, in Appendix E we show that for a prolate we have

$$\lambda + a^2 = \frac{x`^2+y`^2+z`^2-(c^2-a^2)+\sqrt{\{x`^2+y`^2+z`^2-(c^2-a^2)\}^2+4(c^2-a^2)(x`^2+y`^2)}}{2} \quad . \qquad \text{C.15}$$

**Appendix D: Expressing λ In Terms Of The Chebyshev Function $T_{1/3}$.**

Eq. A.4 can be rewritten in the form

$$\lambda^3 + p\lambda^2 + q\lambda + r1 = 0 \quad , \qquad \text{D.1}$$

where $p\,, q$ and $r1$ are given by Eqs. A.9, A.10 and A.11, respectively. With the substitution

$$\lambda = 2\sqrt{\frac{\bar{a}}{3}}\ u - p/3 \quad , \qquad \text{D.2}$$

we obtain, from Eq. D.1,

$$\lambda^3 + p\lambda^2 + q\lambda + r1 = 2\sqrt{\left(\frac{\bar{a}}{3}\right)^3}\ \left[4\,u^3 - 3\,u - \bar{b}\Big/2\sqrt{(\tfrac{\bar{a}}{3})^3}\right] = 0 \quad , \qquad \text{D.3}$$

where $\bar{a}$ and $\bar{b}$ are defined by Eqs. A.7 and A.8 respectively. The Chebechev function of the first kind satisfies

$$4\,(T[\tfrac{1}{3},\omega])^3 - 3T\left[\tfrac{1}{3},\omega\right] - \omega = 0 \quad , \qquad \text{D.4}$$

since the above equation is equivalent to the trigonometric identity

$$4\cos^3\theta - 3\,cos\theta - cos3\theta \equiv 0 \ . \qquad \text{D.5}$$

Therefore , if we identify

$$\omega = \bar{b}\Big/2\sqrt{(\tfrac{\bar{a}}{3})^3} \quad , \qquad \text{D.6}$$

we have $u = T\left[\frac{1}{3},\omega\right]$, and using Eq. D.2,

$$\lambda = 2\sqrt{\frac{\bar{a}}{3}}\ T[\tfrac{1}{3},\omega] - p/3 \quad . \qquad \text{D.7}$$

We can also represent the solution $T[\frac{1}{3},\omega]$ explicitly as a function of $\omega$ as follows:

$$T\left[\tfrac{1}{3},\omega\right] = \left(\tfrac{1}{2}\right)\{\,\left(\omega+\sqrt{\omega^2-1}\right)^{\frac{1}{3}} + (\omega-\sqrt{\omega^2-1})^{1/3}\ \} \ . \qquad \text{D.8}$$

It is interesting to consider the simple case of a sphere where $a = b = c$, so that, from Eqs. A.9, A.10 and A.11, we have

$$p = 3a^2 - r`^2 \quad , \qquad \text{D.9}$$

$$q = 3a^4 - 2a^2 r`^2 \quad , \qquad \text{D.10}$$

$$r1 = a^6 - a^4 r`^2 \quad , \qquad \text{D.11}$$

$$r`^2 = x`^2 + y`^2 + z`^2 \ , \qquad \text{D.12}$$

from which, using Eqs. A.7 and A.8,

$$\bar{a} = (\tfrac{1}{3}) r`^4 \quad , \qquad \text{D.13}$$

$$2\sqrt{\frac{\bar{a}}{3}} = \left(\frac{2}{3}\right) r`^2 \ , \qquad \text{D.14}$$

$$2\sqrt{(\tfrac{\bar{a}}{3})^3} = (\tfrac{2}{27}) r`^6 \quad , \qquad \text{D.15}$$

$$\bar{b} = (\tfrac{2}{27}) r`^6 \quad . \qquad \text{D.16}$$

Therefore $= 1 \, , T\left[\frac{1}{3}, \omega\right] = 1$, and from Eq. D.7,

$$\lambda = \left(\frac{2}{3}\right) r`^2 T\left[\frac{1}{3}, 1\right] - \frac{p}{3} = r`^2 - a^2 \qquad \text{D.17}$$

which also follows trivially from Eq. A.4 with $a = b = c$.

There are two other solutions to Eq. D.3, given by

$$u = \left(\frac{-1}{4}\right)\left\{ \left(\omega + \sqrt{\omega^2 - 1}\right)^{\frac{1}{3}} + \left(\omega - \sqrt{\omega^2 - 1}\right)^{\frac{1}{3}} \right\} \mp \left(\frac{\sqrt{-3}}{4}\right)\left\{ \left(\omega + \sqrt{\omega^2 - 1}\right)^{\frac{1}{3}} - \left(\omega - \sqrt{\omega^2 - 1}\right)^{\frac{1}{3}} \right\} \quad \text{D.18}$$

but these solutions give rise to negative λ ,and therefore are unphysical. For instance, for a spherical solution, we have $\omega = 1$. Hence, from above, $= -1/2$ , and then

$$\lambda = 2\sqrt{\frac{\bar{a}}{3}} \ u - \frac{p}{3} = -(\tfrac{1}{3}) r`^2 - \left(a^2 - \frac{r`^2}{3}\right) = -a^2 \qquad . \qquad \text{D.19}$$

## Appendix E : The Solution λ For A Homogeneous Prolate

With the substitution

$$\tilde{\lambda} = \lambda + a^2 \quad , \qquad \text{E.1}$$

Eq. A.4 becomes

$$\tilde{\lambda}^3 + \tilde{p}\tilde{\lambda}^2 + \tilde{q}\tilde{\lambda} = 0 \quad , \qquad \text{E.2}$$

where

$$\tilde{p} = c^2 - a^2 - r^{`2} \quad , \qquad \text{E.3}$$

and

$$\tilde{q} = -(c^2 - a^2)(x^{`2} + y^{`2}) \quad . \qquad \text{E.4}$$

The solutions of Eq. E.2 are

$$\tilde{\lambda} = 0\ ,\ or\ \lambda = -a^2 \quad , \qquad \text{E.5}$$

$$\tilde{\lambda}(\mp) \equiv \left(\frac{1}{2}\right)(-\tilde{p}\ \mp \sqrt{\tilde{p}^2 - 4\tilde{q}}) \quad , \qquad \text{E.6}$$

with the positive solution $r2 = \tilde{\lambda}(+)$ :

$$\lambda + a^2 = \frac{x^{`2}+y^{`2}+z^{`2}-\left(c^2-a^2\right)+\sqrt{\{x^{`2}+y^{`2}+z^{`2}-(c^2-a^2)\}^2+4(c^2-a^2)(x^{`2}+y^{`2})}}{2} \quad . \qquad \text{E.7}$$

## Appendix F : Demonstrating Eq. (26)

The total constant orbital energy of a test body of mass $m$ and velocity $\vec{V}$ in an elliptical heliocentric trajectory of semi-major axis $a$ is given by (Thornton and Marion, p. 305)

$$-\frac{GmM_\circ}{2a} = \frac{1}{2}mV^2 - \frac{GmM_\circ}{R}. \qquad \text{F.1}$$

When the body distance to the Sun is at aphelion, we have that $R = r_a$, and since

$$2a = r_a + r_p, \qquad \text{F.2}$$

where $r_p$ is the perihelion distance, it follows, using Eq. (F.2) in Eq. (F.1), that

$$r_p = r_a \left[ \frac{V^2}{\frac{2GM_\circ}{r_a} - V^2} \right]. \qquad \text{F.3}$$

Now suppose that $\vec{V}$ represents the heliocentric velocity of the satellite immediately after escaping from its binary orbit, and thus that

$$\vec{V} = \vec{V}_0 + \vec{V}_{Sat}, \qquad \text{F.4}$$

where $\vec{V}_{Sat}$ is the velocity of the satellite relative to the center of mass of the binary system and $\vec{V}_0$ is the velocity of the center of mass relative to the Sun center. The magnitude of $\vec{V}_0$ is determined by applying Eq. (F.1) to the elliptical trajectory, with semi-major axis $a_0$, of the center of mass of the binary system, which tells us that

$$V_0^2 = -\frac{GM_\circ}{a_0} + \frac{2GM_\circ}{r_a} \qquad \text{F.5}$$

Furthermore for this orbit, $\in$, and $a_0$ are related to the aphelion distance by the formula

$$r_a = a_0(1+\in), \qquad \text{F.6}$$

and hence Eq. (A.5) yields

$$V_0^2 = \frac{GM_\circ}{r_a}(1-\in), \qquad \text{F.7}$$

where we are ignoring the very small difference between the aphelion of the satellite and the aphelion of the center of mass of the binary system at the time of the escape. Using Eq. (A.7) and Eq. (A.3) we obtain

$$r_p = r_a \left[ \frac{(1-\in)\frac{V^2}{V_0^2}}{2-(1-\in)\frac{V^2}{V_0^2}} \right]. \qquad \text{F.8}$$

Finally, if $\vec{V}_0$ is in the opposite direction of $\vec{V}_{Sat}$, we have

$$V^2 = (V_0 - V_{Sat})^2, \qquad \text{F.9}$$

and therefore

$$\frac{V^2}{V_0^2} = \left(1 - \frac{V_{Sat}}{V_0}\right)^2 \quad . \qquad \text{F.10}$$

Substituting the above equation in Eq. F.8, we obtain Eq. (26).